\documentclass[preprint,onecolumn,nofootinbib]{revtex4}
\usepackage[colorlinks=true,linkcolor=blue,urlcolor=blue,filecolor=black,citecolor=red,pdfstartview=FitV,pdftitle={},pdfsubject={},pdfkeywords={},pdfpagemode=None,bookmarksopen=true]{hyperref}
\usepackage{graphicx}
\usepackage{amsmath}
\usepackage{amsfonts}
\usepackage{amssymb,ulem}
\usepackage{color,xcolor}%
\usepackage{subfigure}
\usepackage{amsthm,amsmath,amssymb}
\usepackage{mathrsfs}
\usepackage{url}
\usepackage{hyperref}

\usepackage{dcolumn}
\usepackage{float}
\usepackage{appendix}
\begin{document}
\title{Diagnosing quantum phase transitions via holographic entanglement entropy at finite temperature}

\author{Huajie Gong$^{1}$}
\thanks{huajiegong@qq.com}
\author{Guoyang Fu$^{1}$}
\thanks{FuguoyangEDU@163.com}
\author{Peng Liu $^{2}$}
\thanks{phylp@email.jnu.edu.cn}
\author{Chongye Chen $^{2}$}
\thanks{cychen@stu2022.jnu.edu.cn}
\author{Xiao-Mei Kuang$^{1}$}
\thanks{xmeikuang@yzu.edu.cn}
\author{Jian-Pin Wu$^{1}$}
\thanks{jianpinwu@yzu.edu.cn, corresponding author}
\affiliation{
	$^1$ Center for Gravitation and Cosmology, College of Physical Science and Technology, Yangzhou University, Yangzhou 225009, China }
\affiliation{$^2$ Department of Physics and Siyuan Laboratory, Jinan University, Guangzhou 510632, P.R. China}

\begin{abstract}
	
We investigate the behavior of the holographic entanglement entropy (HEE) in proximity to the quantum critical points (QCPs) of the metal-insulator transition (MIT) in the Einstein-Maxwell-dilaton-axions (EMDA) model. Since both the metallic phase and the insulating phase are characterized by distinct IR geometries, we used to expect that the HEE itself characterizes the QCPs. This expectation is validated for certain cases, however, we make a noteworthy observation: for a specific scenario where $-1<\gamma \leq -1/3$, with $\gamma$ as a coupling parameter, it is not the HEE itself but rather the second-order derivative of HEE with respect to the lattice wave number that effectively characterizes quantum phase transitions (QPTs). This distinction arises due to the influence of thermal effects. These findings present novel insights into the interplay between HEE and QPTs in the context of the MIT, and have significant implications for studying QPTs at finite temperatures.

\end{abstract}

\maketitle
\tableofcontents

\section{Introduction}

Quantum phase transitions (QPTs) usually involve strongly correlated electron system which is difficult to quantify \cite{Sachdev:2000}.
As a non-perturbative method, holography builds a bridge between the strongly correlated system and the weakly coupled classical gravitational theory in the large N limit \cite{Maldacena:1997re,Gubser:1998bc,Witten:1998qj,Aharony:1999ti}, which is usually solvable. We can construct a gravitational dual model by holography to attack these strongly correlated problems and address the associated mechanisms of QPTs \cite{Zaanen:2015oix,Hartnoll:2016apf,Baggioli:2022pyb,Baggioli:2019rrs}. As a prominent example of QPTs, the metal-insulator transition (MIT) has been implemented in holographic framework \cite{Donos:2012js,Donos:2014uba,Fu:2022qtz,Donos:2014oha,Donos:2013eha,Ling:2014saa,Baggioli:2014roa,Kiritsis:2015oxa,Ling:2015epa,Ling:2015exa,Ling:2016dck,Mefford:2014gia,Baggioli:2016oju,Andrade:2017ghg,Bi:2021maw} and the associated mechanism has also been addressed that the holographic MIT essentially can be depicted by geometry \cite{Donos:2012js}.
Usually, there are two ways of implementing holographic MIT \cite{Donos:2012js}. One is the infrared (IR) instability induced by the lattice operator, the other is the strength of lattice deformation that induces some kind of bifurcating solutions.

On the other hand, quantum entanglement has been playing an increasingly prominent role in the fields of condensed matter theory, quantum information, black hole physics, and so on. A good measure of quantum entanglement is the entanglement entropy (EE). The counterpart of EE in holography, dubbed as holographic entanglement entropy (HEE), has a simple geometric description that EE for a subregion on the dual boundary is proportional to the area of the minimal surface in the bulk geometry \cite{Ryu:2006bv,Takayanagi:2012kg,Lewkowycz:2013nqa,Hubeny:2007xt,Dong:2016hjy}. It has been shown that HEE can diagnose QPTs and thermodynamic phase transitions \cite{Nishioka:2006gr,Klebanov:2007ws,Jokela:2020wgs,Pakman:2008ui,Zhang:2016rcm,Zeng:2016fsb,Ling:2015dma,Ling:2016wyr,Ling:2016dck,Kuang:2014kha,Guo:2019vni,Mahapatra:2019uql,Liu:2020blk,Baggioli:2023ynu}. Particularly, it has been found that the HEE itself, or its derivatives with respect to system parameters exhibits extremal behavior near quantum critical points (QCPs) \cite{Ling:2015dma,Ling:2016wyr,Ling:2016dck}.

In this paper, we intend to further understand the relation between the EE and QPTs. In \cite{Donos:2014uba}, the authors proposed a special Einstein-Maxwell-dilaton-axions (EMDA) model, for which the spatial linear dependent axion fields couple with a dilaton field\footnote{For a detailed discussion of simple axion models, please consult the original paper \cite{Andrade:2013gsa} and the comprehensive review \cite{Baggioli:2021xuv}.
}. What is vitally important is that the IR geometries of this EMDA model can be analytically expressed such that at zero temperature limit, the scaling behavior of the direct current (DC) resistivity and the low-frequency alternating current (AC) conductivity can be worked out \cite{Donos:2014uba}. This model exhibits rich and meaningful phase structures which is addressed in \cite{Donos:2014uba,Fu:2022qtz}. Particularly, a novel holographic quantum phase transition from a normal metallic phase ($\partial_{T}\sigma_{DC}<0$) with AdS$_2\times \mathbb{R}_2$ IR geometry to a novel metallic one ($\partial_{T}\sigma_{DC}>0$) with non-AdS$_2\times \mathbb{R}_2$ IR geometry was found in \cite{Fu:2022qtz}. The features of their low-frequency AC conductivity indicate that the normal metallic phase behaves as a coherent system while the novel metallic phase exhibits incoherent behavior \cite{Fu:2022jqn}. Further, it is also found that the butterfly velocity or its first derivative exhibiting local extreme behaviors \cite{Fu:2022jqn,Baggioli:2018afg}. In addition, the scaling behavior of the butterfly velocity in the zero-temperature limit confirm that different phases are controlled by different IR geometries \cite{Fu:2022jqn}. Therefore, it is exciting that this EMDA model is able to address so many important issues in the holography community and it is also expected to provide a good platform to attack the aforementioned problem, i.e., the relation between the EE and QPTs. In principle, we should carry out our study at very low temperature as addressed in previous works \cite{Ling:2015dma,Ling:2016wyr,Ling:2016dck}, however, it is extremely difficult to study QPTs within the low-temperature regime of the EMDA model due to the numerical challenges. Thus, here we focus on examining the MIT at finite temperatures, which is of particular importance for practical applications since all real-world systems operate at non-zero temperatures.

The organization of the paper is as follows: Section \ref{sec-EMDA-model} provides a concise introduction to the special EMDA model, highlighting its key features and presenting the corresponding phase diagrams. In Sec.\ref{HEE}, we work out HEE and study the relation between HEE and QPTs. Finally, Section \ref{sec-conclusion} contains the conclusions and discussions.

\section{Holographic background and phase structure}\label{sec-EMDA-model}

The EMDA theory we consider takes the action \cite{Donos:2014uba,Fu:2022qtz}
\begin{eqnarray}\label{EMDA-Action}
	S= {}\int d^{4}x \sqrt{-g} \left[ R +6 \cosh\psi - \frac{3}{2} [ (\partial \psi)^2+4\sinh^2\psi (\partial{\chi})^2 ] - \frac{1}{4} \cosh^{\gamma /3}(3\psi)F^2 \right]\,,
\end{eqnarray}
where $F$ is the Maxwell field defined by $F=dA$, $\chi$ is the axion field, and $\psi$ is the dilaton field coupled with $F$ and $\chi$. $\gamma$ is the coupling parameter, depending on which, the system exhibits rich phase structures as illustrated in Refs.\cite{Donos:2014uba,Fu:2022qtz}. Before proceeding, we would like to emphasize that the holographic Q-lattice models \cite{Donos:2013eha} are equivalent to the EMDA model, featuring a precise coupling between the axion kinetic term and the dilaton. For a detailed discussion, we recommend referring to \cite{Donos:2014uba}.

We assume the following background ansatz:
\begin{eqnarray} \label{metric}
	ds^2&&=\frac{1}{z^2}\big{[}-(1-z)p(z)U(z)dt^2+\frac{dz^2}{(1-z)p(z)U(z)}+V_1(z) dx^2+V_2(z) dy^2\big{]}, \\ \nonumber
	A&&=\mu(1-z)a(z) dt,  \\ \nonumber
	\psi&&=z^{3-\triangle}\phi(z), \\ \nonumber
	\chi&&=\hat{k} x,
\end{eqnarray}
where $p(z)=1+z+z^2-\mu^2 z^3/4$. $\triangle$ is the conformal dimension of the dilaton field $\psi$. In the theory \eqref{EMDA-Action}, it is easy to conclude that $\triangle=2$. As Ref.\cite{Fu:2022qtz}, here we focus on the anisotropic background that the axion field $\chi$ only depends on the $x$-direction of the dual boundary field theory, for which $\hat{k}$ characterizes the lattice wave number.  In our convention, $z=1$ and $z=0$ denotes the locations of the black hole horizon and AdS boundary, respectively. The system \eqref{EMDA-Action} with the ansatz \eqref{metric} can be depicted by four second order ordinary differential equations (ODEs) for $V_1\,, V_2\,, a\,, \phi$ and one first order ODE for $U$. To preserve the asymptotic AdS$_4$ on the conformal boundary ($z=0$), we need impose the following boundary conditions:
\begin{eqnarray}
	U(0)=1\,, \ V_1(0)=1\,, \ V_2(0)=1\,, \ a(0)=1\,, \ \phi(0)=\hat{\lambda}\,,
	\label{bc-UV}
\end{eqnarray}
where $\hat{\lambda}$ is the source of the dilaton field operator in the dual boundary field theory and depicts the strength of lattice deformation. Then, we impose the regular boundary conditions at the horizon ($z=1$). Further, we have the Hawking temperature:
\begin{eqnarray}
	\hat{T}=\frac{12-\mu^2}{16 \pi}\,,
	\label{H-T}
\end{eqnarray}
where we have set the boundary condition as $U(1)=1$.
We focus on the canonical ensemble and set the chemical potential $\mu$ as the scaling unit. Thus, for given parameter $\gamma$, this system is completely described by the three dimensionless parameters $\{T , \lambda , k \} \equiv \{\hat{T}/\mu , \hat{\lambda}/\mu , \hat{k}/\mu \}$.

When $\chi=\psi=0$, the background solution \eqref{metric} reduces to the $Reissner-Nordstr\ddot{o}m$ anti-de Sitter black hole (RN-AdS black hole), whose IR geometry is AdS$_2\times \mathbb{R}_2$. By studying the perturbations about this IR fixed point, we can obtain the scaling dimension of the dilaton field operator as
\begin{eqnarray} \label{delta_psi}
	\delta_{+}^{\psi}=-\frac{1}{2}+\frac{1}{6}\sqrt{24 e^{-2 v_{10}} k^2-3(12\gamma+1)}\,,
\end{eqnarray}
where $v_{10}$ can be determined by the IR datas \cite{Donos:2014uba,Fu:2022qtz,Fu:2022jqn}.
When the scaling dimension satisfies $\delta_{+}^{\psi}\geq 0$, which gives
\begin{eqnarray}
	\label{relation-gamma-k}
	2e^{-2 v_{10}}k^2\geq 1+3\gamma\,,
\end{eqnarray}
the IR solution is always RG stable. In addition, it is found that when the lattice wave number $k$ vanishes, i.e., $k=0$, the scaling dimension $\delta_{+}^{\psi}$ is minimized. Based on the above observation, this system can be classified into the following three cases in terms of the parameter $\gamma$ \cite{Donos:2014uba,Fu:2022qtz}:
\begin{itemize}
	\item \textbf{Case I:} $-1<\gamma \leq -1/3$
	
	In this case, the relation $\delta_{+}^{\psi}>0$ always holds at $k\neq 0$, which suggests an irrelevant deformation in IR. That is to say, the IR geometry is RG stable.
	
	\item \textbf{Case II:} $-1/3<\gamma \leq -1/12$
	
	When $\gamma$ in the region of $-1/3<\gamma \leq -1/12$, $\delta_{+}^{\psi}<0$ at $k=0$, which indicates the IR solution to be RG unstable. Further, if the lattice wave number is turned on, i.e., $k\neq 0$, the IR solution can be also RG unstable when the relation \eqref{relation-gamma-k} is violated as the case of $k=0$. Therefore, when reducing $k$ or increasing $\lambda$, one has a RG unstable IR solution, which drives a MIT \cite{Donos:2014uba}.
	
	\item \textbf{Case III}: $\gamma>-1/12$
	
	When $\gamma>-1/12$, $\delta_{+}^{\psi}$ becomes complex at $k=0$. It means that the BF bound is violated resulting in a dynamical instability, which induces a novel black hole with scalar hair. Depending on the parameter $\gamma$, this novel black hole has different ground states at zero temperature that it is insulating for $-1/12<\gamma<3$, and metallic for $\gamma>3$, which can be determined by the DC and AC conductivities over the IR fixed point \cite{Donos:2014uba}.
\end{itemize}

\begin{figure}[H]
	\centering
	\includegraphics[width=0.4\textwidth]{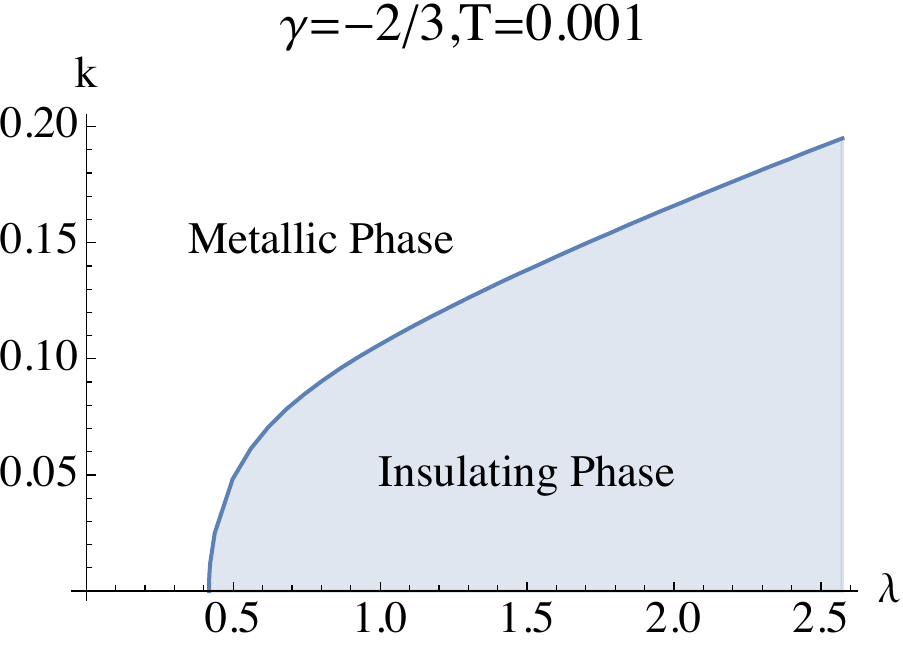}\hspace{0.4mm}
	\includegraphics[width=0.4\textwidth]{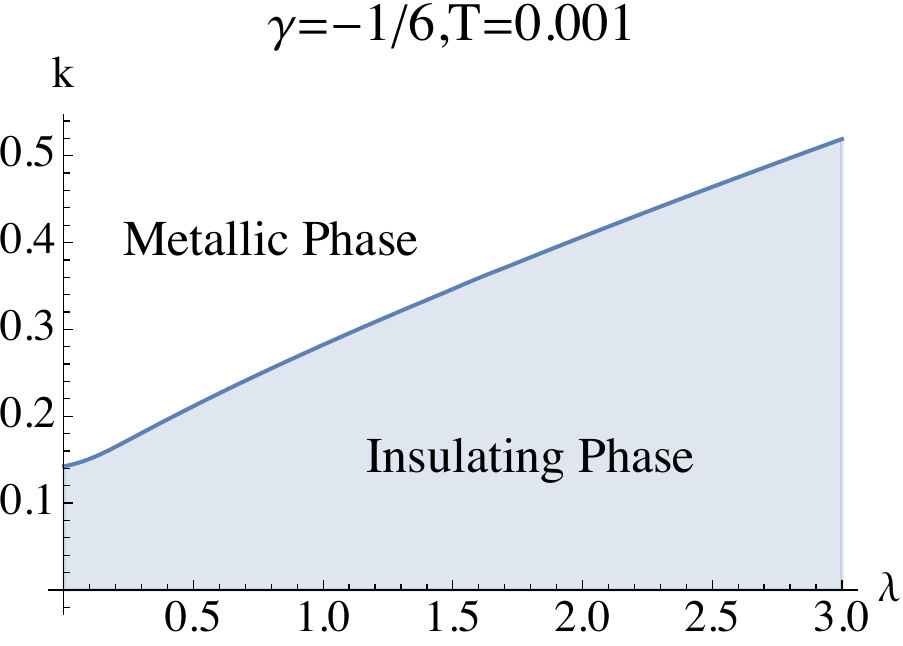}\vspace{0.4mm}
	\includegraphics[width=0.4\textwidth]{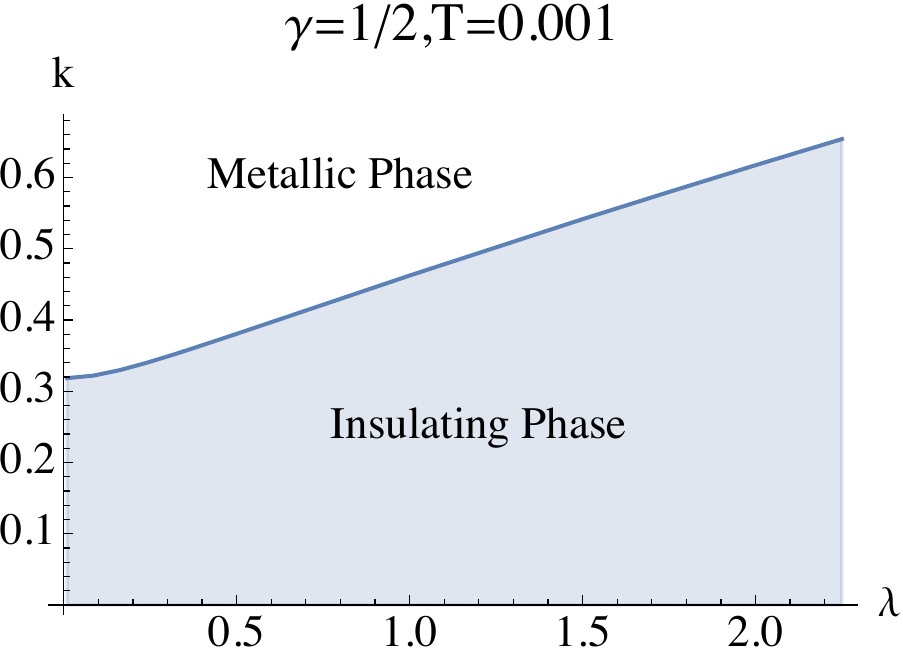}\hspace{0.4mm}
	\includegraphics[width=0.4\textwidth]{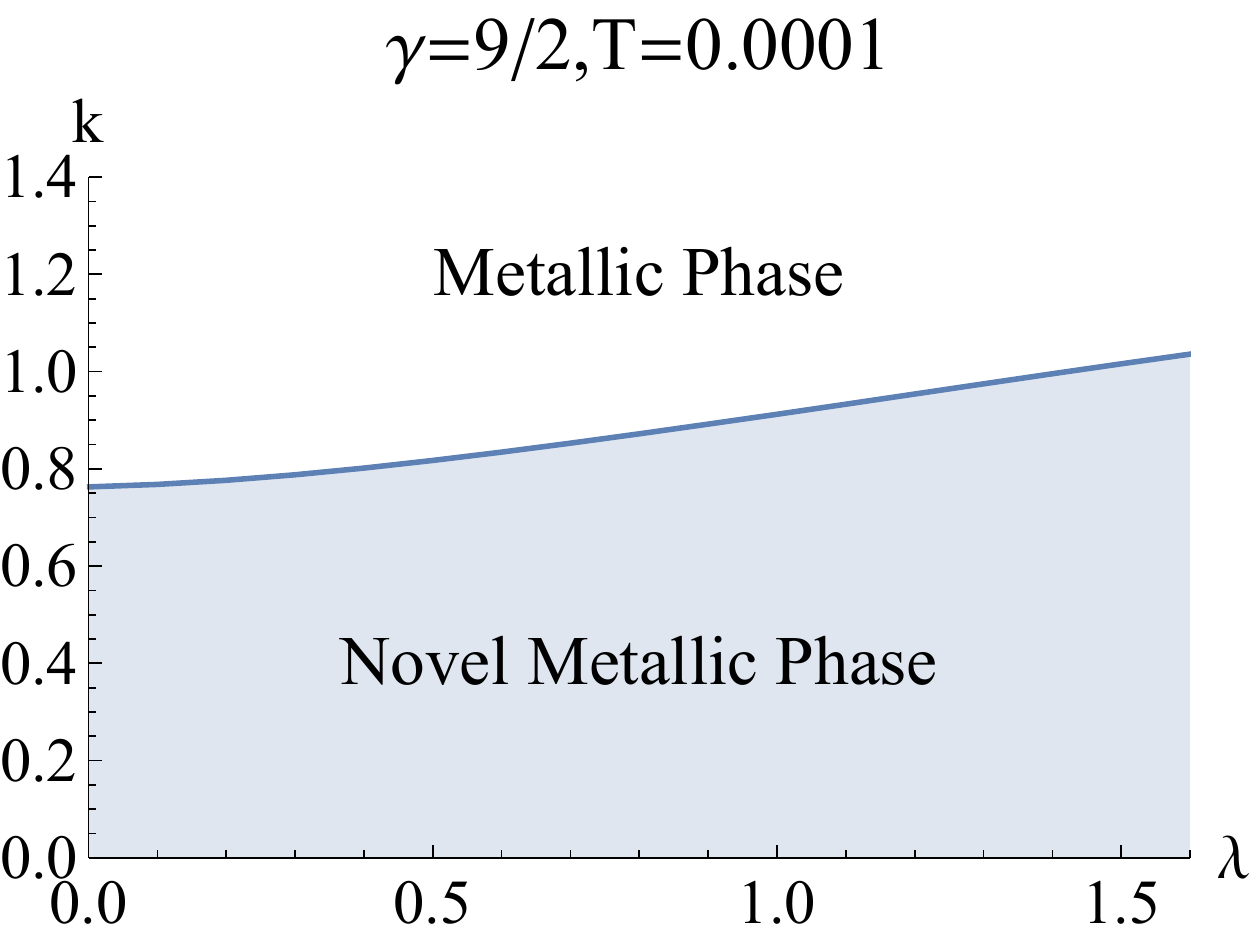}\hspace{0.4mm}
	\caption{The phase diagram over the $\{\lambda, k\}$ with different $\gamma$. The blue line is the critical line of the phase transition.}
	\label{phase diagram}
\end{figure}

To distinguish between metallic and insulating phases, we employ a commonly accepted operational definition, as elucidated in several holographic references \cite{Baggioli:2016rdj,Donos:2012js,Donos:2013eha,Donos:2014uba,Ling:2015ghh,Ling:2015dma,Ling:2015epa,Ling:2015exa,Ling:2016wyr,Ling:2016dck,Baggioli:2014roa,Baggioli:2016oqk,Baggioli:2016oju,Donos:2014oha,Liu:2021stu,Kiritsis:2015oxa,Liu:2022gme,Li:2022yad,Bai:2023use}. 
Specifically, we can determine whether the system is in a metallic or insulating phase based on the sign of $\partial_{T}\sigma_{xx}$. A negative value indicates a metallic phase, while a positive value indicates an insulating phase.
Therefore, in terms of the temperature behaviors of DC conductivity at extremely low temperatures, we can numerically work out the phase diagram over $\lambda$ and $k$ (Fig.\ref{phase diagram}). Though the IR geometry is RG stable for Case I, we still observe a MIT emerging (the upper left plot in Fig.\ref{phase diagram}). It can be attributed to the existence of bifurcating solutions as argued in \cite{Donos:2012js}. We would like to point out that thanks to the AdS$_2\times \mathbb{R}_2$ IR geometry at zero temperature, when the strength of the lattice $\lambda$ is small, the phase is metallic even for small $k$. Different from the Case I, the MIT always exists for any $\lambda$ in Case II (the upper right plot in Fig.\ref{phase diagram}). It is because there is a transition from the AdS$_2\times \mathbb{R}_2$ IR fixed point to a non-AdS$_2\times \mathbb{R}_2$ IR fixed point when enhancing $\lambda$ or reducing $k$, which induces a RG relevant lattice deformation. This mechanism is just the one of the original Q-lattice models studied in \cite{Donos:2013eha}. While for Case III, since the system has different ground states depending on the parameter $\gamma$, it exhibits completely different phase structures (see the bottom plots in Fig.\ref{phase diagram}). For $\gamma=1/2$, the system exhibits an insulating ground state, and correspondingly there is a MIT when reducing $k$. The phase diagram is very similar to that of Case II (the bottom left plot in Fig.\ref{phase diagram}). For $\gamma=9/2$, the system has a metallic ground state. The IR fixed point of this metallic phase is a non-AdS$_2\times \mathbb{R}_2$ geometry, and thus we call it as a novel metallic phase. The phase transition happens from the novel metallic phase to the normal metallic phase with AdS$_2\times \mathbb{R}_2$ geometry when we increases $k$ at fixed $\lambda$ to exceed some critical values (see the right plot in Fig.\ref{phase diagram}). For more detailed discussions on the phase structures, please refer to \cite{Fu:2022qtz}.

\section{Holographic entanglement entropy near QCP}\label{HEE}

The HEE can be computed using the so-called Ryu-Takayanagi (RT) formula \cite{Ryu:2006bv,Takayanagi:2012kg,Lewkowycz:2013nqa}\footnote{The RT formula is reformulated as the Hubeny-Rangamani-Takayanagi (HRT) formula for covariant cases \cite{Hubeny:2007xt,Dong:2016hjy}.}:
\begin{eqnarray}\label{RT_formula}
	S_A=\frac{Area(\gamma_A)}{4G_N}\,,
\end{eqnarray}
where $G_N$ is the bulk Newton constant and $\gamma_A$ is the minimal surface extending from the boundary subregion $A$ into the bulk. Without loss of generality, we investigate simply an infinite strip subsystem in the dual boundary, which can be formally characterized as $A:=\{0 < x < l, -\infty < y < \infty\}$.
We may explicitly write out the HEE and the associated width of the strip for the EMD-axions model investigated here:
\begin{eqnarray}
	&&
	\label{Expression-of-HEE}
	\hat{S}= 2\int_{0}^{z_*} \left[ \frac{z_*^2\sqrt{V1(z)}V2(z)}{z^2 \sqrt{G(z)}\sqrt{z_*^4 V1(z) V2(z) - z^4 V1(z_*) V2(z_*)}}-\frac{1}{z^2 }\right]dz-\frac{2}{z_*}\,,
	\
	\\
	&&
	\label{Expression-of-width}
	\hat{l}= 2\int_{0}^{z_*} \left[ \frac{z^2 \sqrt{V1(z_*)V2(z_*)}}{\sqrt{G(z) V1(z)}\sqrt{z_*^4 V1(z) V2(z) - z^4 V1(z_*) V2(z_*)}} \right]dz\,,
\end{eqnarray}
where $G(z)=(1-z)p(z)U(z)$. Here, a counterterm $-1/z^2$ has been inserted to cancel out the vacuum contribution. $z_*$ indicates turning point of the minimal surface along the $z$-direction. In what follows, we will primarily focus on the scaling-invariant HEE and width, denoted by $S\equiv\hat{S}/\mu$ and $l\equiv\hat{l}\mu$, respectively.
In this section, we will study the characteristics of HEE near QCPs over this EMDA model.

\begin{figure}[H]
	\centering
	\includegraphics[width=0.45\textwidth]{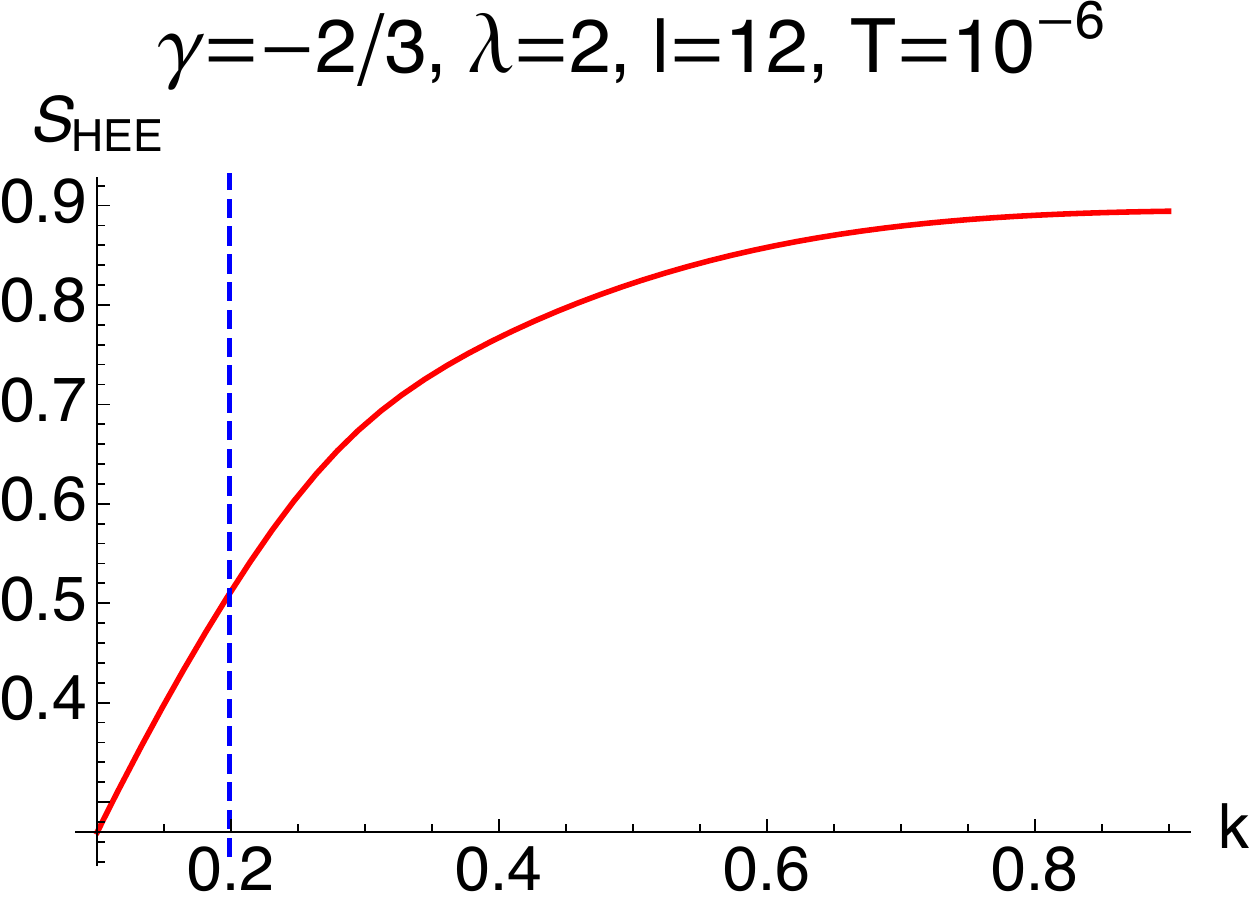}\hspace{10mm}	
	\includegraphics[width=0.45\textwidth]{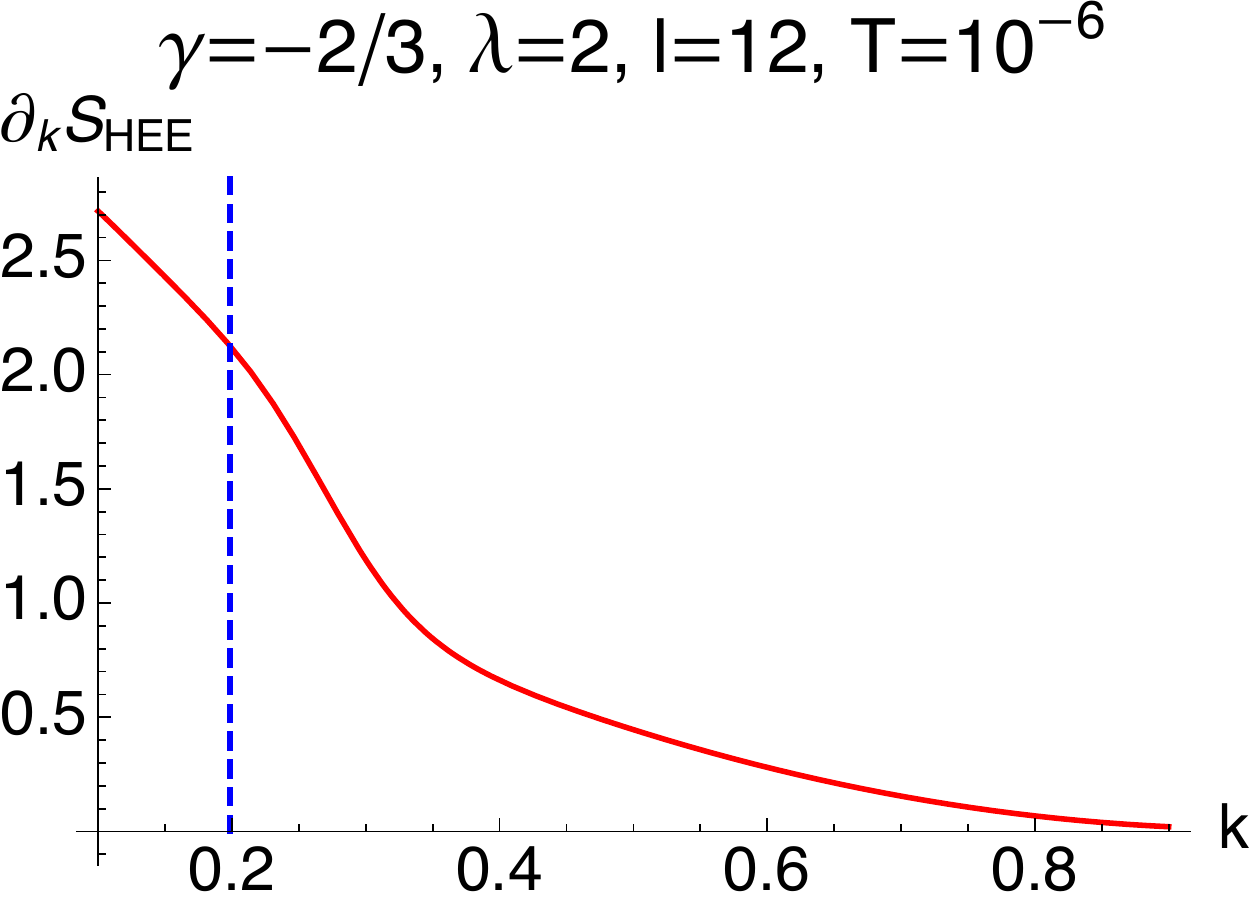}
	\caption{The HEE $S_{HEE}$ and its first-order derivative with respect to $k$ as a function of $k$ for $\gamma=-2/3$ at $T=10^{-6}$. The blue dashed line is the position of the QCP. Here we have set $\lambda=2$ and $l=12$.
}
	\label{HEEvsk-caseI}
\end{figure}

\begin{figure}[H]
	\centering
	\includegraphics[width=0.4\textwidth]{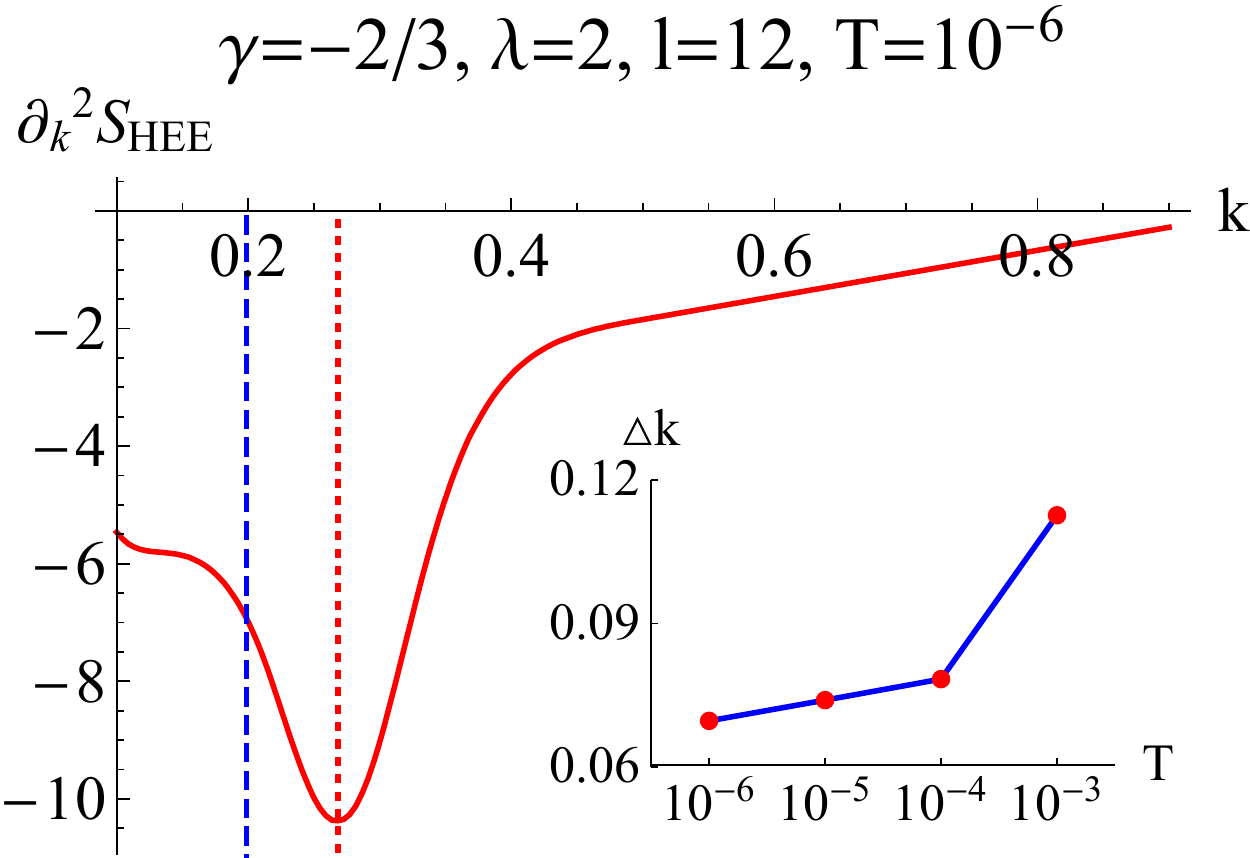}\hspace{10mm}	
	\includegraphics[width=0.5\textwidth]{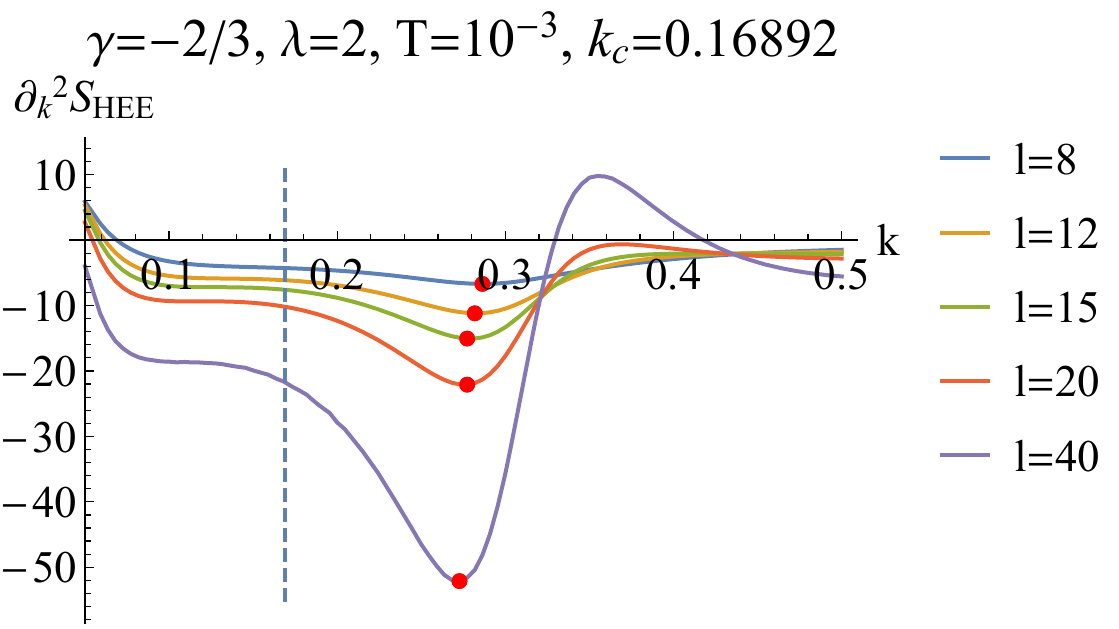}
	\caption{$\partial^2_k S_{HEE}$ as a function of $k$. In left plot we have fixed $l=12$ at extremal low temperature $T=10^{-6}$, while in right plot we fix the temperature $T=10^{-3}$ with different $l$. The blue dashed line is position of QCPs. The red dashed line (left plot) or the red points (right plot) denote the local minimum of $\partial^2_k S_{HEE}$. The insets in the left and right plots display $\Delta k$ as a function of $T$ and $l$, respectively.
}
	\label{HEE2vsk-caseI}
\end{figure}

We firstly explore the behavior of HEE for Case I. We would like to emphasize that the MIT in this scenario cannot be induced by the IR geometry instability because the IR geometry is RG stable \cite{Donos:2014uba,Fu:2022qtz}. A possible mechanism is that the MIT is driven by the strength of the lattice deformation, which results in the bifurcating solutions \cite{Donos:2012js}. Without loss of generality, we choose $\gamma=-2/3$ and an extreme low temperature $T=10^{-6}$. Fig.\ref{HEEvsk-caseI} illustrates the HEE itself and its first-order derivative with respect to $k$, i.e., $\partial_k S_{HEE}$ as a function of $k$. In this scenario, neither the HEE nor its first-order derivative displays extremal or singular behavior near QCPs. In contrast, HEE goes up and its first-order derivative goes down monotonically with $k$, even when the system changes from the insulating phase to the metallic one. We observe, however, that when transitioning from the insulating phase to the metallic one, $\partial_k S_{HEE}$ exhibits a significant reduction of orders of magnitude (right plot in Fig.\ref{HEEvsk-caseI}). Based on this observation, it is expected that the QCPs can be captured by the local extreme of the second-order derivative of HEE, i.e., $\partial^2_k S_{HEE}$. So, we further show $\partial^2_k S_{HEE}$ as a function of $k$ in Fig.\ref{HEE2vsk-caseI}. The left plot of this figure reveals that the local minimum of $\partial^2_k S_{HEE}$ is located relatively close to the QCPs, validating our inference. Also, we use the symbol $\Delta k$ to represent the difference between the location of the QCPs and the local minimum of $\partial^2_k S_{HEE}$, as illustrated in the inset of the left plot in Fig.\ref{HEE2vsk-caseI}. We find that $\Delta k$ goes down monotonically as temperature drops. It indicates that in this case the QPT may be captured by the local extreme of $\partial^2_k S_{HEE}$ in the limit of zero temperature. Additionally, we also show $\partial^2_k S_{HEE}$ as a function of $k$ for various $l$ at $T=10^{-3}$. Notice that at low temperature and large $l$, the numerical calculation becomes more difficult and time consuming. As a result, we fix $T=10^{-3}$ in the right plot of Fig.\ref{HEE2vsk-caseI}. Nevertheless, we still observe that when $l$ increases, the local minimum of $\partial^2_k S_{HEE}$ approaches the QCPs. The inset in this plot further corroborates this observation. It implies that in both limits of large $l$ and zero temperature, the diagnosis of QCPs using the local minimum of $\partial^2_k S_{HEE}$ becomes evident.

\begin{figure}[H]
	\centering
	\includegraphics[width=0.4\textwidth]{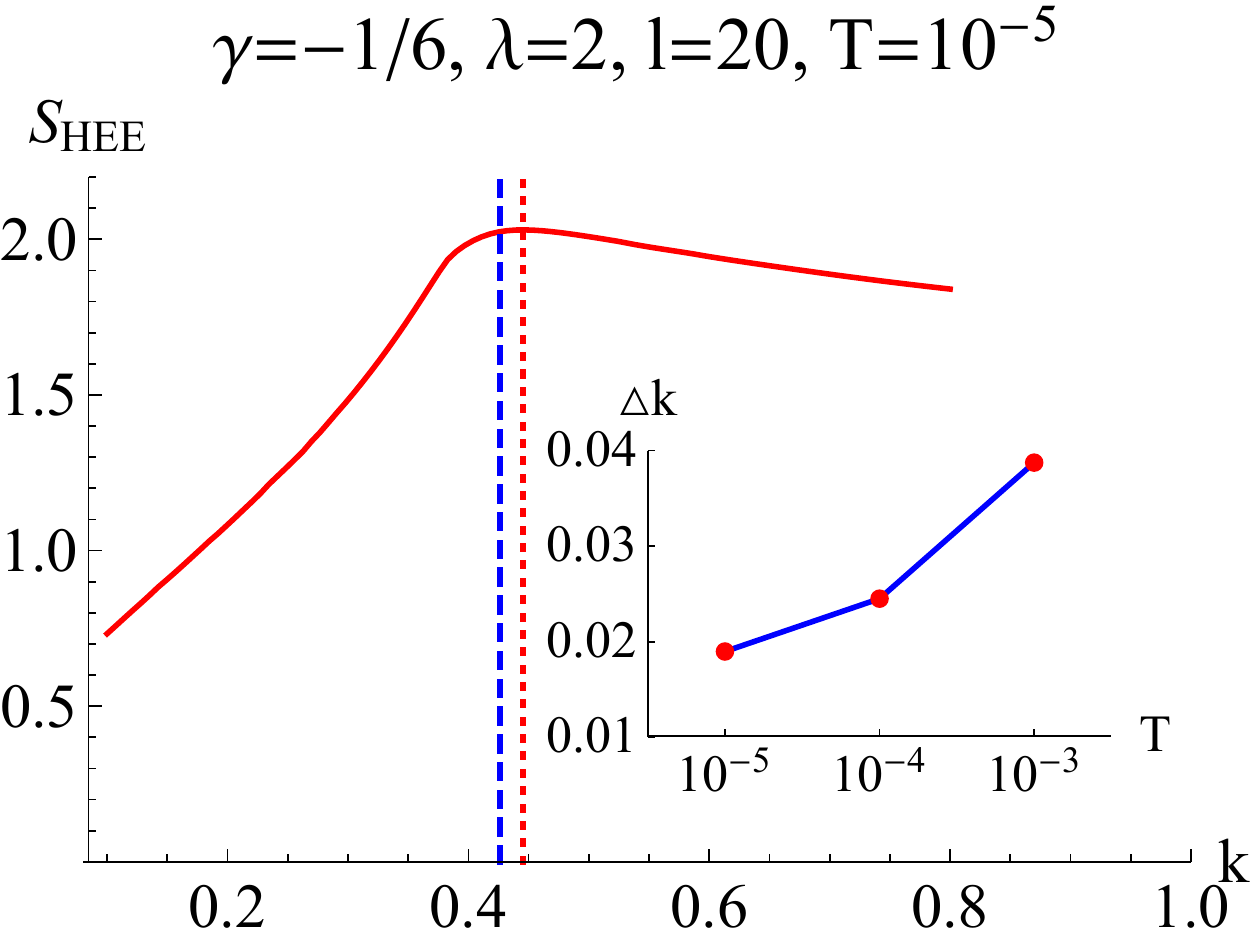}\hspace{10mm}	
	\includegraphics[width=0.5\textwidth]{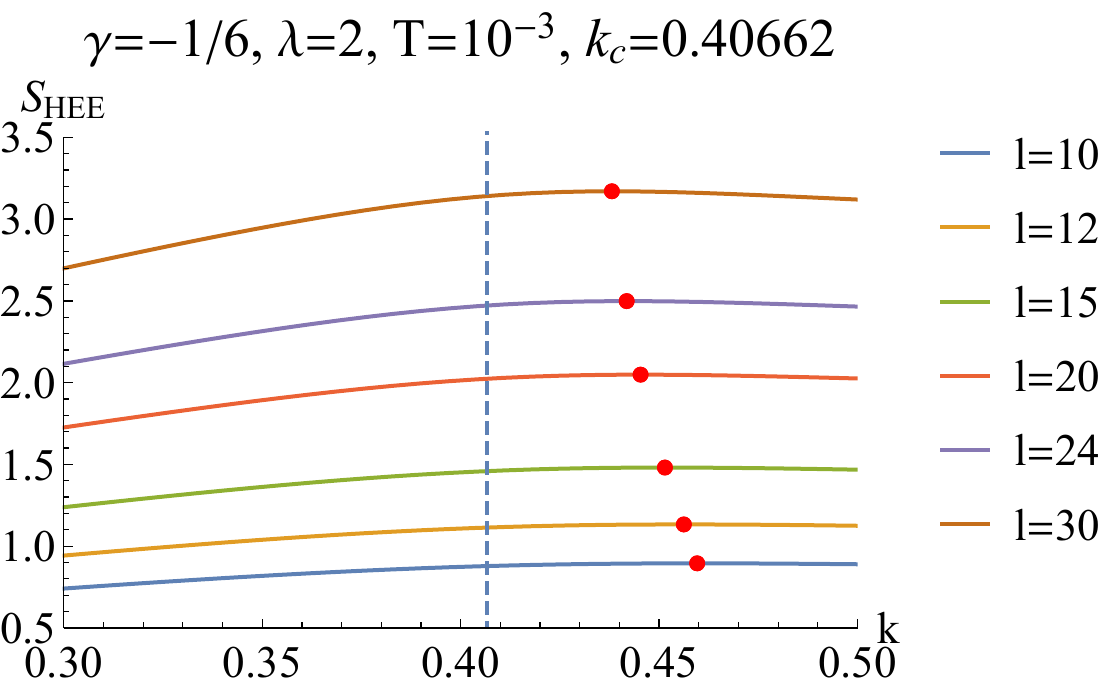}
	\caption{$S_{HEE}$ as a function of $k$. In left plot we have fixed $l=20$ at extremal low temperature $T=10^{-5}$, while in right plot we fix the temperature $T=10^{-3}$ with different $l$. The blue dashed line is position of QCPs. The red dashed line (left plot) or the red points (right plot) denote the local maxima of $S_{HEE}$. The insets in the left and right plots display $\Delta k$ as a function of $T$ and $l$, respectively.
	}
	\label{HEEvsk-caseII}
\end{figure}

For Case II, the MIT happens due to the IR geometry instability. This mechanism is identical to that of the standard Q-lattice model investigated in \cite{Donos:2013eha}. The left plot in Fig.\ref{HEEvsk-caseII} shows $S_{HEE}$ as a function of $k$ at extremal low temperature $T=10^{-6}$. Here we have fixed $\gamma=-1/6$ and $l=20$. We observe that HEE itself displays a local maximum, which is different from Case I. Similarly to Case I, we use $\Delta k$ to show the difference between the position of QCPs and the local maximum of $S_{HEE}$, as seen in the inset of the left plot of Fig.\ref{HEEvsk-caseII}. We discover that $\Delta k$ falls monotonically as the temperature drops. Therefore, we conclude that in Case II, the HEE itself is capable of diagnosing the QPT at the limit of zero temperature. This conclusion is compatible with the standard Q-lattice model \cite{Ling:2015dma}. We also show how $S_{HEE}$ changes as a function of $k$ for different $l$ at $T=10^{-3}$ in the right plot of Fig.\ref{HEEvsk-caseII} and $\Delta k$ as a function of $l$ in the inset of this plot. We observe that as $l$ goes up, the local maximum of $S_{HEE}$ gets closer and closer to the QCP. It means that in both limits of large $l$ and zero temperature, the diagnosis of the QCP using the local maximum of $S_{HEE}$ becomes evident.

\begin{figure}[H]
	\centering
	\includegraphics[width=0.4\textwidth]{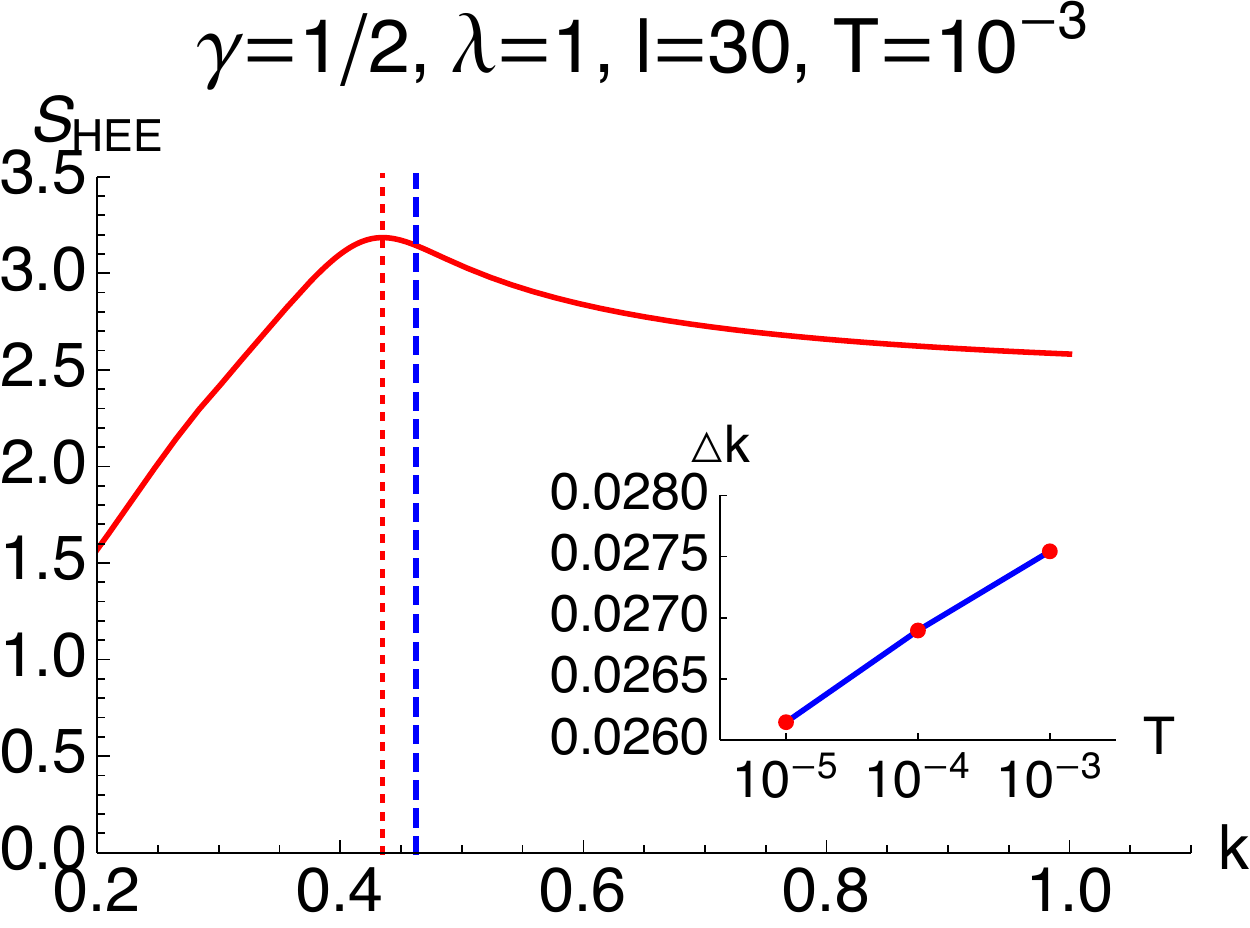}\hspace{10mm}	
	\includegraphics[width=0.5\textwidth]{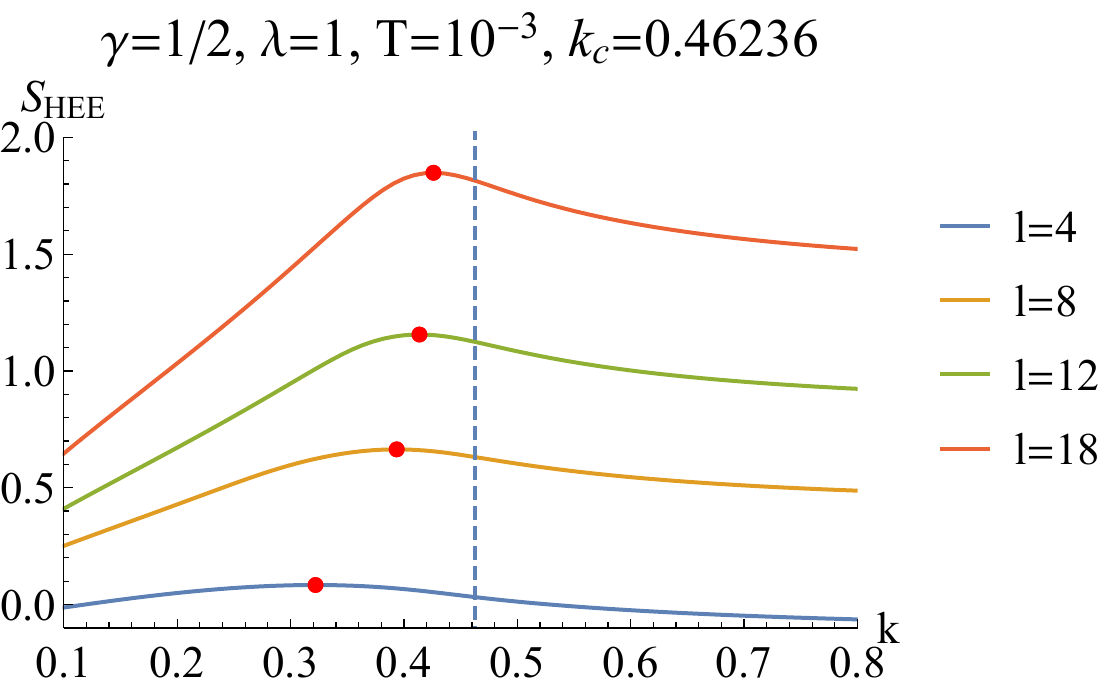}\ \\
	\includegraphics[width=0.4\textwidth]{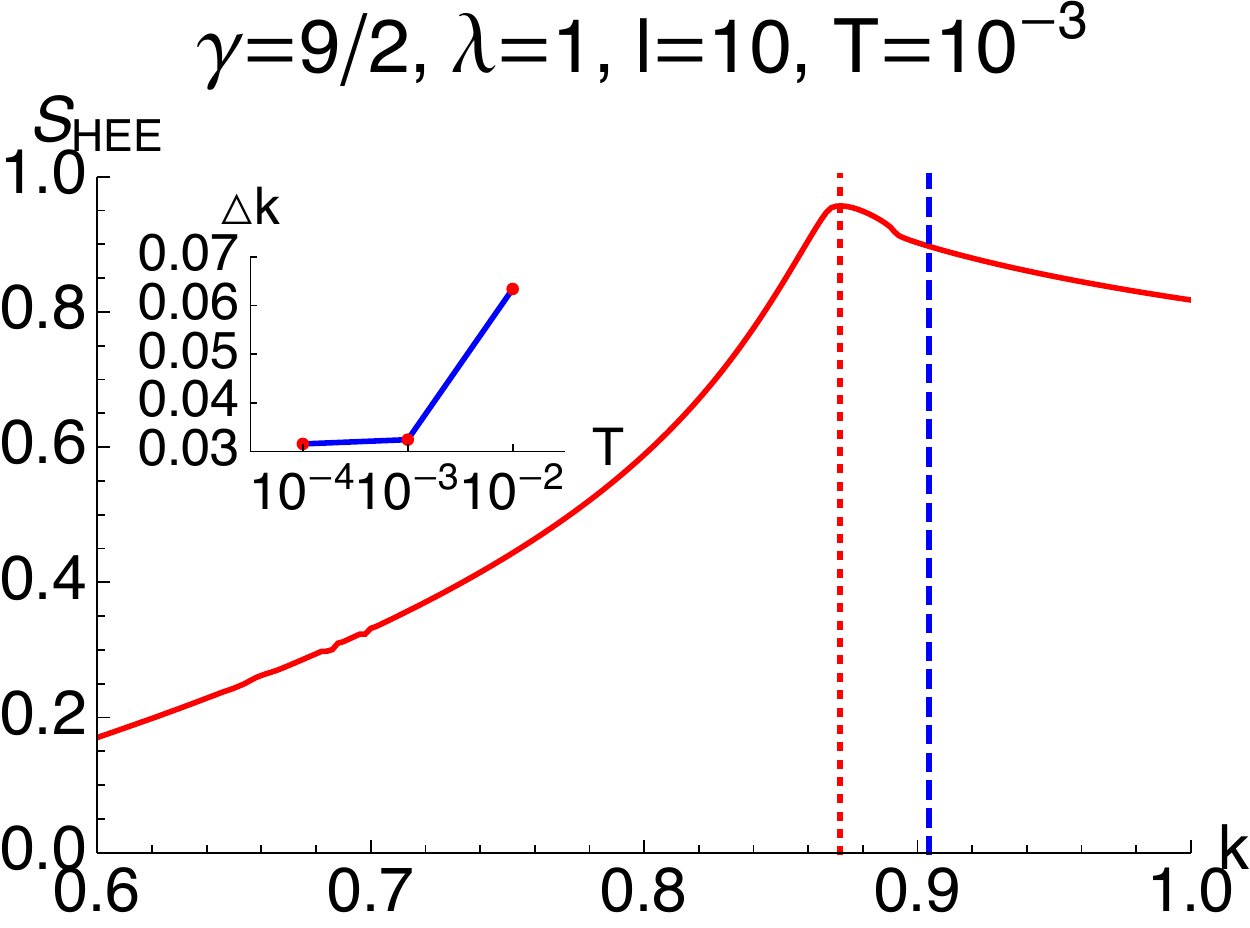}\hspace{10mm}	
	\includegraphics[width=0.5\textwidth]{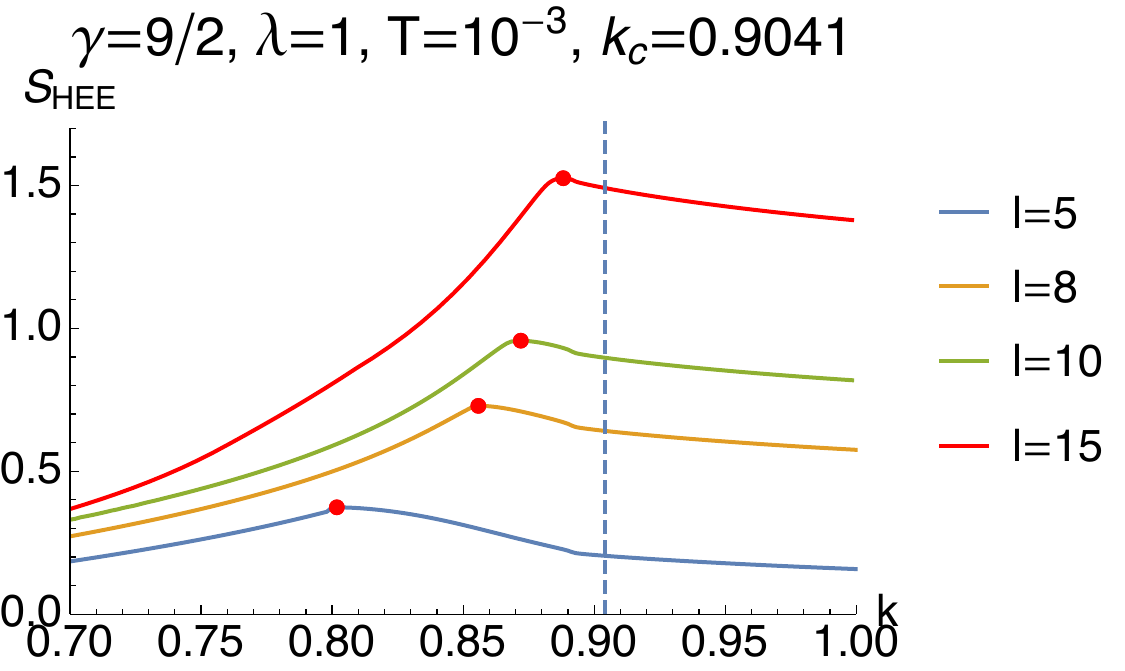}
	\caption{$S_{HEE}$ as a function of $k$ for Case III. In left plot we have fixed $l=30$ for $\gamma=1/2$ and $l=10$ for $\gamma=9/2$ at $T=10^{-3}$, while in right plot we fix the temperature $T=10^{-3}$ with various $l$. The blue dashed line is position of the QCP. The red dashed line (left plot) or the red points (right plot) denote the local maxima of $S_{HEE}$. The insets in the left and right plots display $\Delta k$ as a function of $T$ and $l$, respectively.
	}
	\label{HEEvsk-caseIII}
\end{figure}
Now we come to Case III, where the MIT happens because a novel black hole with scalar hair develops when we change $\lambda$ or $k$. This novel black hole exhibits different ground states at zero temperature depending on the parameter $\gamma$. Fig.\ref{HEEvsk-caseIII} shows the HEE behaviors at $\gamma=1/2$ and $\gamma=9/2$, where the ground state is insulating and metallic, respectively. We find that the HEE almost exhibits the same behaviors as Case II. That is to say, the HEE itself is capable of diagnosing the QPT at the limit of zero temperature.

\section{Conclusion and discussion}\label{sec-conclusion}

This paper builds upon previous investigations into the relationship between HEE and QPTs. In our series of studies, we have made several key observations:
\begin{enumerate}
	\item In \cite{Ling:2015dma}, we study the HEE behavior of holographic Q-lattice model. In this model, the metallic phase displays an AdS$_2 \times \mathbb R^2$ IR geometry at zero temperature, resulting in a non-vanishing ground state entropy density, whereas the insulating phase showcases a hyperscaling violation IR geometry at zero temperature with a vanishing ground state entropy density. Our findings reveal that the HEE exhibits local extremes in the vicinity of the QCPs of the MIT at extremely low temperature. This is expected due to the noticeable differences in the IR geometry between the metallic phase and the insulating phase.
	\item In Gubser-Rocha model with Q-lattices \cite{Ling:2016wyr}, both the metallic and insulating phases manifest hyperscaling violation IR geometry at zero temperature, leading to a vanishing ground state entropy density. In this case, the difference between the IR fixed points for metallic phase and insulating phase are less significant, as a result, diagnosing the QCPs solely using the HEE itself becomes challenging in this scenario. However, our findings reveal that it is the first-order derivative of the HEE with respect to the system parameter that effectively diagnoses the QCPs in the MIT. Our study provides compelling evidence that HEE can still effectively detect QPTs in these circumstances, suggesting its potential for broader and more realistic applications in quantum many-body systems.
	\item In our study, presented in \cite{Ling:2016dck}, we further examine a holographic axion model incorporating a non-minimal coupling between the matter field and the gravity theory. We discover that this model also displays a MIT. For both the metallic and insulating phases of this model, the IR geometry manifests as AdS$_2$ at zero temperature, resulting in an identical non-vanishing ground state entropy density. We found that in this model, the second order derivative of HEE with respect to the axionic charge can effectively characterize the QPT. It is because the non-minimal coupling between the matter field and the gravity theory can modify the prescription of HEE, meaning that the matter field can influence HEE, thereby reflecting the QCPs, despite the geometry itself being no difference from AdS$_2 \times \mathbb R^2$.
\end{enumerate}

In comparison to our previous work, the IR geometry of the EMDA model is AdS$_2 \times \mathbb R^2$ in the metallic phase, while it displays hyperscaling violation features in the insulating phase. Notably, the characteristics of the IR geometry of the EMDA model studied here shares similarities with the original holographic Q-lattice model \cite{Donos:2013eha}. Therefore, it is expected that HEE itself can diagnose the QCPs, similar to the finding in \cite{Ling:2016wyr}.
In cases II and III, we have confirmed that HEE characterizes the QCPs, as expected. However, for Case I, it is the second order derivative of HEE with respect to the lattice wave number that characterizes the QPT, but not the HEE itself. This distinction can be attributed to the influence of thermal effects. In Case I, at low temperatures, finding the solutions of the minimum surfaces and their resultant HEE becomes numerically challenging, so our study was limited to higher temperature. At higher temperature, the signatures of the QCPs are potentially buried by thermal effects, making it difficult to diagnose them. Nonetheless, we show that even at finite temperature, HEE can still reflect the QCPs, albeit through their second order derivatives, rather than HEE itself. This finding is of particular significance as it pertains to real-world systems, which are inherently finite temperature. By leveraging this approach, we can gain deeper insights into the underlying physics governing the QPT in quantum many-body system.

	Finially, we would like to offer some insights regarding the use of EE as a diagnostic tool for identifying QCPs, although this remains a subject of speculation. Following the pioneering work by \cite{2002Natur.416..608O}, numerous studies, such as \cite{Lambert:2004zz,Latorre:2004pk,2006NJPh....8...97C,PhysRevB.73.224414}, have aimed to demonstrate that entanglement-related measures or their derivatives, can be employed in QCP diagnosis. It has been observed that entanglement measures can serve as indicators of phase transitions. Sometimes, the entanglement measures themselves are sufficient to diagnose these transitions, while in other cases, it is necessary to consider their derivatives. This distinction arises from the fact that any entanglement measure can be expressed as certain functional of the set of first derivatives of the ground state energy \cite{PhysRevA.74.052335}. Therefore, in order to unveil the underlying phase transitions, it becomes crucial to analyze the derivatives of these entanglement measures. Holographic duality offers new insights into the connection between entanglement and QPTs. In cases where two IR fixed points exhibit significant differences, such as one having zero ground state entropy while the other does not, the entanglement itself can distinguish between the two phases \cite{Ling:2015dma}. However, when the differences between two different IR fixed points are relatively less significant, it may be necessary to consider derivatives to reveal the phase transitions \cite{Ling:2016wyr,Ling:2016dck}. This situation is reminiscent of similar cases in condensed matter theory. Furthermore, it is important to note that QPTs are typically described at zero temperature in principle. However, both realistic systems and holographic dualities impose limitations, confining us to the finite temperature regime. In this scenario, the entanglement measures alone may not be sufficient to distinguish between the two phases due to the significant thermal fluctuations present. These fluctuations can overshadow the quantum entanglement, rendering it as subleading terms. Nevertheless, the derivatives of the entanglement measures can still exhibit strong signals near the critical points, enabling them to effectively reveal the occurrence of phase transitions.

Our investigation of the EMDA model has yielded valuable insights into the connection between HEE and the MIT at finite temperature, and has opened up several exciting future avenues for research. One such direction would be to assess other models including QPTs at relatively higher temperature and examine whether taking higher derivatives of HEE could expose the QCPs even further. This potentially opens up an new area of research into QPTs at finite temperature. Moreover, this work provides a useful tool to identify QCPs in cases where locating them is a challenge. Analysis of second-order or even higher order derivatives can provide a signal to detect phase transitions at numerically accessible regions, making it easier to locate the critical points before digging deeper into the more time-consuming low-temperature details. This approach offers the potential to extend our understanding of phase transitions across more general models, opening up new avenues of QPTs research in the holographic framework and even in the quantum many-body system.

It is intriguing to delve deeper into the behaviors of various information measures, such as holographic mutual information, the holographic entanglement of purification, and the c-function, in the vicinity of holographic QCPs. One notable example is the holographic mutual information and the holographic entanglement of purification, which have been shown to be effective probes for studying thermal phase transitions \cite{Liu:2020blk}. In addition, the study in \cite{Baggioli:2020cld} has revealed that the c-function can serve as a novel and accurate probe for detecting the location of topological QCPs. 

\acknowledgments
This work is supported by the Natural Science Foundation of China under Grant No. 11905083, No. 12375054 and No. 12375055, the Science and Technology Planning Project of Guangzhou (202201010655), Natural Science Foundation of Jiangsu Province under Grant No.BK20211601, Fok Ying Tung Education Foundation under Grant No.171006, and the Postgraduate Research $\&$ Practice Innovation Program of Jiangsu Province under Grant No.KYCX22$\_$3451.
Peng Liu would like to thank Yun-Ha Zha for her kind encouragement during this work.

\bibliographystyle{style1}
\bibliography{Ref}
\end{document}